\documentclass[12pt]{article}
\begin{document}                                                     

\renewcommand{\theequation}{\thesection.\arabic{equation}}        
\newcommand{\mysection}[1]{\setcounter{equation}{0}\section{#1}}

\def\1{{\bf 1}}
\def\Z{{\bf Z}}
\def\ee{\end{equation}}
\def\be{\begin{equation}}
\def\l{\label}
\def\dxi{\partial_\xi}
\def\D{{\cal D}}
\def\sin{{\rm sin}}
\def\cos{{\rm cos}}
\def\f{{\bf \Phi}}
\def\v{\varphi}
\def\O{\bf {\cal O}}
\def\C{\bf C}
\def\CP{\bf CP}
\def\e{\rm e}
\def\0{\nonumber}
\def\eea{\end{eqnarray}}
\def\bea{\begin{eqnarray}}
\def\Tr{\rm Tr}
\def\IR{\bf R}
\def\ZZ{\bf Z}
\def\T{\tau}

\begin{titlepage}

\hfill{ULB-TH/02-30}

\vspace{2 cm}

\centerline{{\huge{Matrix String Theory on pp-waves}}}

\vspace{2 cm}

\centerline{Giulio Bonelli\footnote{e-mail address: gbonelli@ulb.ac.be}}
\centerline{Physique Theorique et Mathematique}
\centerline{Universite' Libre de Bruxelles}
\centerline{Campus Plaine C.P. 231; B 1050 Bruxelles, Belgium}
\centerline{and}
\centerline{Theoretische Natuurkunde}
\centerline{Vrije Universiteit Brussel}
\centerline{Pleinlaan, 2; B 1050 Brussel, Belgium}

\vspace{4 cm}

{\it Abstract}: After a breif review on Matrix String Theory on flat backgrounds,
we formulate matrix string models on different pp-wave backgrounds.
This will be done both in the case of constant and variable RR background
flux for certain exact string geometries.
We exhibit the non--perturbative representation of string interaction 
and show how the eigenvalue tunneling drives the WKB expansion to give
the usual perturbative string interaction also in supersymmetric 
pp-wave background cases.

\vfill
{\it Proceedings of the RTN-Workshop
``The quantum structure of spacetime and the geometric nature of fundamental interactions'',
Leuven, September 2002}
\end{titlepage}

\paragraph{Overview of Matrix String Theory}

Matrix String Theory \cite{etc} (MST) connects perturbative string theory
and M(atrix)-theory \cite{bfss} by explaining how the theory of strings is included 
in the latter in a second quantization scheme. 
As known, eleven dimensional M-theory compactified on a circle $S^1_{(11)}$ is
supposed to give rise to the type IIA string theory at strong coupling
where D-particles are claimed to be the natural degrees of freedom 
describing the theory.
To be able to reach a string theory in a perturbative regime, one may
be able to use a strong/weak duality, namely the type IIB S-selfduality.
Therefore, in order to perform such a program, one is forced to further
compactify the theory on another circle $S^1_{(9)}$ to be able to reach
type IIB string theory by a $T_{(9)}$-duality.
Insodoing, the D-particles get mapped to D-strings whose winding in the 
nineth direction counts their number.
Now it is possible to perform an S-selfduality of the type IIB string theory
by mapping these D-strings in fundamental type IIB strings 
with the same winding in the nineth direction and weak coupling.
Upon a final $T_{(9)}$-duality these strings get mapped to weakly coupled
type IIA strings with fixed light cone momentum.
Notice that we started with a M-theory  picture in which $R_{(11)}=g_s l_s$
is a much bigger circle than $R_{(9)}$  -- being the type IIA theory in the strong coupling
regime -- while we end up in an M-theory picture in which
$R_{(11)}=g_s l_s$ is a much smaller than $R_{(9)}$.
Actually, the chain of dualities we performed could have been obtained 
simply as a role-flip in M-theory between the two circles.
The outcome of the above duality chain is that 
perturbative type IIA strings admit a dual description as wrapped type IIB D-strings
once the light-cone momentum of the strings is associated to the winding
number of the D-strings.
Let us notice that from this point of view the overall transverse spacial eight
dimensional geometry plays a passive role and that it can be left as an
arbitrary string background.
Anyway, the need of tractable cases tends to simplify the transverse geometry
in a natural way. Therefore, we review the flat case.

The effective description of wrapped type IIB D-strings is obtained by
considering the massless spectrum of the open strings stretched between them.
This is given by the dimensional reduction to the D-string worldsheet 
of the $D=1+9$ {\cal N}=1 superYang-Mills theory with gauge group $U(N)$, that is
the $D=1+1$ {\cal N}=(8,8) superYang-Mills theory with gauge group $U(N)$.
The gauge theory coupling $g$ weights the mass scale of the stretched open
strings, that is the D-string tension, and therefore it is natural to set
$g=\frac{1}{l_sg_s}$ from the duality chain, where $g_s$ is the resulting type
IIA perturbative string coupling. 
If all that is true, the 
superconformal fixed point of the (8,8) superYang-Mills theory with gauge group
U(N) at the strong Yang-Mills coupling limit and in the large N regime
has to realizes the free limit of type IIA string theory on flat background.
To show that this is indeed the case,
let us consider for simplicity just the bosonic part of the action first, that is
$$
S_b=\frac{1}{2}\int d^2z\left\{
D_{\bar z}X^I D_zX^I+
\frac{1}{2g^2}|F_{z\bar z}|^2
-\frac{g^2}{2}\sum_{IJ}[X^I,X^J]^2
\right\}
$$ 
where the $X^I$ are in the $8_v$ of the R-symmetry $SO(8)$ group
representing the unbroken transverse rotational group
and are in the adjoint representation of the gauge group.
The two dimensional integral is along a cylinder $S^1\times {\bf R}$.
By taking the strong coupling limit we see 
that classically , i.e. analyzing the potential
$V=\frac{g^2}{2}\sum_{IJ}[X^I,X^J]^2$, the  
degrees of freedom surviving the strong coupling limit are
the Cartan projected matrix fields such that $[X^I,X^J]=0$.
The same argument applies to the gauge connection, once the gauge coupling
gets rescaled in the covariant derivatives and in the gauge curvature.
Therefore the $X^I$ are simultaneusly diagonalizable by a large gauge
transformation, i.e. $X^I=U x^I U^\dagger$ where $x^I$ is a diagonal
matrix. Since the only gauge invariant concept is the spectrum of the 
$X^I$s itself, we find that the IR fixed point has to be given by 
a supersymmetric $\sigma$-model on the quotient $\left({\bf R}^8\right)^N/S_N$.
This classical extimation can be made precise by considering
the full theory
\be
S=S_b+S_f\quad
{\rm where}\quad
S_f=\frac{1}{2}\int d^2z\left\{
i\Theta_s D_z \Theta_s + i\Theta_c D_{\bar z} \Theta_c
+2ig \Theta_s\Gamma^I\left[X^I,\Theta_c\right] 
\right\}
\l{fullflat}\ee
where the fermion fields $\Theta_s$ and $\Theta_c$ are in the $8_s$ and $8_c$
representations of $SO(8)$ and in the adjoint of the gauge group.
As the bosons, also the fermions get projected to the Cartan in the strong
coupling.
Expanding the action around a generic Cartan valued field configuration, 
it is easy to see that in the strong coupling limit the only left over
contribution to the integration along the non-Cartan degrees of freedom
is a supersymmetric Gaussian path integral.
This shows the quantum stability of the Cartan projection at strong coupling.
and means that at strong coupling we find an IR dynamics governed by the
action
$$
S^{\infty}=
\frac{1}{2}\int d^2z\left\{
D_{\bar z}x^I D_zx^I+
\frac{1}{2}|f_{z\bar z}|^2+
i\theta_s D_z \theta_s + i\theta_c D_{\bar z} \theta_c
\right\}
$$
where all the fields are in a common Cartan subalgebra.
As it is easy to see, the above action -- due to the fact that the gauge connection
is completly decoupled from the $\sigma$-model fields and that
it does not carry any local degree of freedom --
is equal in form to the GS type IIA
action, but its field content is a bit different being given by a symmetrized
sum of $N$ copies of it, that is an orbifold $\sigma$-model on $\left({\bf R}^8\right)^N/S_N$.
This is because the large gauge transformation $U$ relating
the single valued field $X^I$ and the Cartan projected $x^I$ as
$X^I=U x^I U^\dagger$ is not required to be periodic, namely upon a $2\pi$ shift
in the space circle one has 
$ U\quad\to\quad U\cdot g$ and $x^I\quad\to\quad g^\dagger\cdot
x^I\cdot g$ where $g\in S_N.$
To disentangle the orbifold structure, we fix the conjugacy classes of each
permutation $g\in S_N$ in irreducible cyclic ones as
\be
(g)=(1)^{n_1}\cdot (2)^{n_2}\cdot\dots\cdot(a)^{n_a}\cdot\dots\cdot(N)^{n_N}
\quad{\rm where}\quad
\sum_a an_a=N
\l{conj}\ee
Let us focus on a single cyclic permutation $(a)$.
It acts on the relative eigenvalues of the Cartan fields as
$x^I_\alpha(\sigma+2\pi)=x^I_{\alpha+1}(\sigma)$ where
$\alpha\in[1,\dots,a]$.
Considering the collection of the $x_\alpha^I$s, we can collect them in a
single field $\hat x^I$ with period $2a\pi$.
Therefore the field content of the sector corresponding to the
generic permutation (\ref{conj}) is given by a set of strings
(one for each cyclic elementary permutation)
of length $a$. In the large $N$ limit, this is the representation of the second
quantized GS type IIA string in the free limit, 
where the length parameter is mapped to the light-cone string momentum
\footnote{See \cite{DLect} for an extensive review of this point.}.

As far as the string interaction is concerned, there are two different
appreaches. The first, which has been considered in \cite{etc}, is a
constructive one and goes as follows.
One considers the effect of switching back perturbatively the
gauge coupling. This should result as an additional vertex 
$
S^\infty\quad\to\quad S=S^\infty+\frac{1}{g}\int d^2z V^{(3/2,3/2)}+\dots
$
driving the orbifold CFT out of the conformal point.
By $S_N$ and $SO(8)$ symmetries and locality arguments
the vertex $V^{(3/2,3/2)}$ can be fixed uniquely to be given 
in terms of spin operators. The vertex indeed reproduces the Mandelstam
vertex insertion at the branching points of a typical string diagram.

A second approach to string interactions, which has been considered in
\cite{MST}, starts by looking for the string interaction in the very structure
of the gauge theory.
The first observation is that different long string configurations
corresponds to topologically dinstinct vacua of the gauge theory.
The possibility therefore arises that some instantonic effect takes place by
tunneling between these different vacua.
This is indeed the case.
Non trivial instantons (preserving 1/2 of the original supersymmetry) are
given by field configurations satisfying the Hitchin system
\cite{H}
\be
F_{z\bar z}+g^2[\sigma,\bar\sigma]=0
\quad{\rm and}\quad
D_z\sigma=0
\l{hitchin}\ee
where $\sigma=X^1+iX^2$ and the other scalar fields are passive.
The spectral data classifying its solutions is given by the moduli
space ${\cal H}_N$ of holomorphic plane curves
of rank $N$. 
The generic curve ${\cal S}$ is defined by the $\sigma$-spectral equation
$$
0={\rm det}(\sigma(z)-\sigma{\bf 1}_N)
$$
for any solution of (\ref{hitchin}).
The way in which these spectral curves represent
the Mandelstam diagrams is discussed in detail in \cite{MST}.
The branching points of the curve ${\cal S}$ where various sheets
come together are therefore associated to the joining/splitting 
of strings.
The gauge curvature is generically turned on only in a region 
of size $g^{-1}$ around 
the branching points of the curve ${\cal S}$ where various sheets
come together, while away from these points it is small corresponding to 
the stability of the intermediate free string configurations where
$\sigma$ is a normal matrix.
This string interpretation of the Hitchin system allows then to make manifest
the eigenvalue tunneling responsable for the string interaction
and to build it directly from the gauge theory.
Infact one can refine the strong coupling limit by calculating the WKB 
approximation of a generic matrix string amplitude by expanding the action 
around the solutions of the Hitchin system.
Once a particular solution, i.e. a spectral curve ${\cal S}$, is choosen, then 
it singles out a precise Cartan subalgebra pattern almost everywhere and the fields
eigenvalues now become all together the corresponding light-cone string
coordinate fields on the spectral curve itself.

Therefore we obtain a representation for the partition sum as
$$
Z_N=\sum_{{\cal S}\in{\cal H}_N}\int D[x,\theta] e^{-S(x,\theta)_{\cal S}}\times 
\int D[A] e^{-S(A)_{\cal S}}
$$
where $S(x,\theta)_{\cal S}$ is the type IIA GS string action on ${\cal S}$
and $S(A)_{\cal S}$ is the action of a decoupled U(1) gauge theory on ${\cal S}$.
The integration over the gauge field is trivial and gives
$$\int D[A] e^{-S(A)_{\cal S}}=\frac{Det'\Delta_0}{Det'\Delta_1} 
g^{\P_0-\P_1}$$
where $\Delta_i$ are the Laplacian for $i$-differentials on ${\cal S}$
and $\P_i$ is the number of their zero modes.
By using the Riemann-Roch formula $\P_0-\P_1=\chi_{\cal S}$ we can rewrite
the matrix string partition sum as
\footnote{We redefine the measure on ${\cal H}_N$ to include also 
$\frac{Det'\Delta_0}{Det'\Delta_1}$.}
\be
Z_N=\sum_{{\cal S}\in{\cal H}_N}
g_s^{-\chi_{\cal S}}
\int D[x,\theta] e^{-S(x,\theta)_{\cal S}}
\l{mmspf}\ee
where we identify (after the proper introduction of dimensionfull constants)
$g=\frac{1}{l_sg_s}$.
Eq.(\ref{mmspf}) shows that the two dimensional gauge theory defining 
matrix string theory admits an asymptotic expansion as a GS string theory
already at finite $N$. To match the usual GS partition function
we have to make sure that the moduli space of the Hitchin systems ${\cal H}_N$
recostructs the full moduli space of Riemann surfaces.
This can be shown to happen exactly in the large $N$ limit.
As far as the interpretation of the resulting measure as the Weil-Petersson
in the large $N$ limit, see \cite{Brax}.

This way, the interacting structure of perturbative string theory, 
namely the genus expansion, is recovered as an asymptotic expansion of 
the gauge theory partition function around the 
conformal fixed point in the inverse gauge coupling which is then,
accordingly with S-duality, interpreted as the string coupling.

\paragraph{Strings and pp-waves}

While it is still an open issue the quantization of string theory in curved backgrounds
and the study of their finiteness properties,
it appeared recently an interesting class of (2,2) pp-wave solutions of type IIB 
\cite{mama} generalizing the one studied in \cite{metsaev}.
These string theories have been shown to admit a supercovariant 
formulation in \cite{bema} and have been shown to be exact (finite)
in \cite{bema,ruts}.
These (2,2) solutions have been studied in the $SU(4)\times U(1)$
formalism \cite{GS}.
Within this framework the type IIB GS action on a flat background 
is written in terms of four complex chiral (2,2) superfields 
$X^{+l}$, 
where $l=1,\dots,4$, as
$$
S_0=\int d^2z d^4\theta X^{+l}X^{-l}
$$
where the only part of the original $SO(8)$ symmetry which remains manifest is 
a $SU(4)\times U(1)$ one.
The action relative to these new pp-wave backgrounds is written as the $\sigma$-model action
$$
S^{IIB}_{pp}=\int d^2z \left\{\int d^4\theta X^{+l}X^{-l}
+\int d^2\theta W(X^{+l}) + c.c.\right\}
$$
where $W$ is the prepotential, i.e. an holomorphic function in the four
chiral fuperfields. The pp-wave metric and the RR-field
curvature are parametrized by this holomorphic function as
$
ds^2_{(10)}=-2dx^+dx^--|\partial W|^2(dx^+)^2+2dx^l d\bar x^l
$
and the RR-fields $F^{(5)}\sim \partial^2 W$.
Notice that from
this exact superconformal formulation, upon
a change of variables, the R-NS formulation can be obtained where the string
interaction is well defined, being the background non dilatonic, 
as the usual genus expansion.
Particularly, if $W(x)=\mu\sum_l x_l^2$, then we obtain the maximal
supersymmetric pp-wave with manifest $SO(4)$ symmetry and constant
RR-flux $F^{(5)}\sim\mu$ while in general
if $W$ is a generic quadratic function, these backgrounds have
constant RR-flux and less isometries.

To formulate a Matrix String Theory for such a kind of backgrounds
(in order to embed them in M-theory) we have to consider their type IIA
counteparts.
Because of the different chirality assignment between left and right moving 
fermions, the analogous type IIA (2,2) $\sigma$-model has just three
chiral superfields $\phi^i$,
where $i=1,2,3$, and a twisted chiral
\footnote{For definitions, properties and notation here and in the subsequent part of this letter,
we refer the reader to \cite{notation}.}
one that we call $\Sigma$.
This is the type IIA counterpart of the above superfields formalism in which
only an $SU(3)\times U(1)$ symmetry is manifest.
The type IIA action is then given by
\be
S^{IIA}_{pp}=\int d^2z \left\{\int d^4\theta \left\{-\frac{1}{4}\Sigma\bar\Sigma+ \phi^i\bar\phi^i
\right\} +\int d^2\theta W(\phi^i) + c.c.\right\}
\l{22GSA}\ee
where the prepotential depends now only upon the three chiral superfields.
In principle one could also consider more general $\sigma$-model actions
including also generalized FI terms like 
$\int d^2z \left\{\int d\theta^+d\bar\theta^- f(\Sigma) +c.c.\right\}$, where
$f$ is holomorphic, but we don't do it here. 
Notice that, in particular, if $W=0$, this action reproduces the type IIA GS action on a flat background.
Now the possibility of lifting to M-theory (via a MST picture) these backgrounds
can be posed clearly.

\paragraph{Matrix Strings on pp-waves with constant RR flux}

As far as the formulation of MSTs on pp-waves with {\it constant} RR-flux
backgrounds, the procedure to obtain the most general kind of model
is to consider a generic deformation of the MST action (\ref{fullflat}) which is quadratic in
the fermions. 
This is because we already know that the $\sigma$-model remains quadratic in
the fermion fields.
This procedure can be best studied in the ten dimensional formalism where
the fermion fields $\Theta_s$ and $\Theta_c$ are collected in a single
Weyl-Majorana fermion we call $\Psi$.
Infact type IIA Matrix String Theory can be obtained by reducing the 
${\cal N}=1$ D=10 SYM theory with gauge group $U(N)$ down to two dimensions
and we can therefore work directly in the ten dimensional theory
to calculate the constraints on the possible quadratic deformations 
and then dimensionally reduce.
This procedure has been worked out in \cite{ppmst}.
Summarizing, we deform the ${\cal N}=1$ D=10 SYM action
$$
S_{sym}\quad\to\quad S_{pp}=
S_{sym}+ \int i\bar\Psi H\Psi +S_b
$$
where $H$ is a constant bi-spinor whose generic structure
is $H=h_{\mu\nu\rho}\Gamma^{\mu\nu\rho}\left(\frac{1+\Gamma^{11}}{2}\right)$
and $S_b$ is a purely bosonic action term. This has to be fixed by the 
requirement that a shifted supersymmetry
$$
\delta A_\mu = \delta_{sym} A_\mu
\quad
\delta \Psi = \delta_{sym}\Psi + K[A]\epsilon
$$
leaves invariant the shifted action.
This is possible if $K[A]$ is a linear functional in the gauge field 
$A$ and if the supersymmetry parameters $\epsilon$ satisfy
certain linear differential equations.
The action for the general matrix string theory model on pp-waves with
constant RR flux can be obtained by dimensional reduction to two dimensions 
of the above construction.
The details and the results of this approach can be found in \cite{ppmst} 
and will not be given here.

\paragraph{Matrix Strings on (2,2) pp-waves with non-constant RR flux}

This second case is the implementation in the Matrix String Theory
scheme of the exact (2,2) pp-wave backgrounds reviewed in the last section
\cite{iopp}. In order to treat this case, we preliminarly
write down the MST action on flat background in the (2,2) superfield formalism.
It reads as
\be
\int d^2z d^4\theta \left(-\frac{1}{4g^2}\Sigma\bar\Sigma + \Phi^i\bar\Phi^i\right)
+ \int d^2z \left[d^2\theta L(\Phi^i) + c.c.\right]
\l{22mst}
\ee
with the prepotential $L=g \Phi^1\left[\Phi^2,\Phi^3\right]$.
Here $\Phi^i$ are three chiral superfields in the adjoint 
representation of the gauge group $U(N)$ while $\Sigma$
is the twisted chiral superfield obtained as the covariant
superfield curvature. The trace over the gauge group and the covariantization
of the $\Phi\bar\Phi$ term are understood in (\ref{22mst}).
Notice that in the (2,2) manifest formalism only a $U(1)\times SU(3)$
subgroup of the original $SO(8)$ R-symmetry is manifest (exactly as it was 
in the previous section for the action (\ref{22GSA}) in the flat case.).

Now, to include the $\sigma$-model prepotential $W$, we add a further
prepotential term in (\ref{22mst}) as
\be
S=\int d^2z \left\{\int d^4\theta \left\{-\frac{1}{4g^2}\Sigma\bar\Sigma+  
\Phi^i\bar\Phi^i
\right\} +\int d^2\theta \left(L(\Phi^i)+\tilde W(\Phi^i)\right) + c.c.\right\}
\l{pp22mst}\ee
As far as the definitions of the matrix function $\tilde W$
the natural requirement is that once evaluated on Cartan fields it reproduces
the prepotential in (\ref{22GSA}) as 
$\Tr\tilde W\left({\Phi^t}^i\right)=\sum_m
W\left({\Phi_m}^i\right)$ in an ortonormal basis of $t$.
This requirement specifies these structure function up to matrix ordering.
The natural ordering is of course the total symmetrization.
In the related context of D-geometry this problem has been 
clarified \cite{dgeom} by showing that if the background satisfies the string
equations, then the total symmetric ordering is the correct one
in order to reproduce the correct open string masses assignments.
Since the string backgrounds we are considering are exact, 
TS-duality with the type IIB D-string picture justifies this ordering.
Moreover, there's a second argument in favour of the symmetric ordering which
is related to the symmetries of the background.
Our type IIA generic background is expilicitly invariant under $SO(2)$ (acting on the 
$\sigma$ complex plane) and the $SU(3)$ transformations under which 
$W$ is invariant (up to additional constants).
Let us notice that, since the additional prepotential $L=g\Phi^1[\Phi^2,\Phi^3]$
is fully $SU(3)$ invariant and since the total symmetrization 
is the only matrix ordering prescription which commutes with linear
transformations of the arguments, then this prescription
is the only one which preserves the above background symmetries.
Let us notice that this agrees with the issue raised in \cite{rudi}
where the symmetry of the background pp-wave metric is token
as a guiding principle for the construction of a well defined 
string perturbation theory.
We consider these arguments exaustive of a discussion about the matrix
ordering prescription.

For completeness, we give the bosonic part of the action (\ref{pp22mst})
that is
$$
S_b=\frac{1}{2}\int d^2z\left\{
D_{\bar z}\phi^iD_z\bar\phi^{\bar i}+
D_{z}\phi^iD_{\bar z}\bar\phi^{\bar i}+
D_{\bar z}\sigma D_z\bar\sigma+
D_{z}\sigma D_{\bar z}\bar\sigma+
\right.$$ $$\left.
\bar F^{\bar i}F^i+\frac{1}{2g^2}|F_{z\bar z}|^2
-\frac{g^2}{2}[\sigma,\bar\sigma]^2-\frac{g^2}{2}[\phi^i,\bar\phi^{\bar i}]^2
+g^2[\bar\sigma,\phi^i][\sigma,\bar\phi^{\bar i}]
+g^2[\sigma,\phi^i][\bar\sigma,\bar\phi^{\bar i}]
\right\}
$$ 
where 
$\bar F^i=\frac{\partial(\tilde W+L)}{\partial\Phi^i}(\phi)$.

A first check of our model is that in the strong coupling it has to 
reduce to a symmetrized orbifold of the type IIA GS action (\ref{22GSA}).
This is indeed the case, since the additional prepotential $\tilde W$ is
independent upon the gauge coupling and therefore, assuming that the RR flux
is weak with respect to the gauge coupling $g$, it does not interfere with
the strong coupling limit procedure as described for the flat case in the 
first section.

A second important point is to check if the WKB expantion of the matrix string
partition function still reproduces the perturbative expansion of string
theory in a way similar to the one we revied in the first section.
This can be again shown to happen in this case (the case of generic
constant RR flux \cite{ppmst} seems to be more difficult because of lack of control on
the supersymmetric instantons if any).
Actually, the possible BPS field configurations for the MST on pp-wave we are
studying are of two types.
\begin{itemize}
\item{
Static: these are 1/2 BPS solutions where the $\sigma$ field is passive,
the gauge connection is flat and the $\phi^i$s satisfy the equations
$$
D_0\phi^i=0
\quad
D_1\phi^i+ F^i=0
\quad
\sum_i \left[\phi^i,\bar\phi^i\right]=0
$$}
\item{
Instanton: again 1/2 BPS solutions where the $\phi^i$s are passive and 
$\sigma$ and the gauge connection satisfy again the Hitchin equations
(\ref{hitchin}).
}
\end{itemize}
Since $F$-term/derivative-terms saturation implies staticity, 
there does not exist other istanton type solutions but the ones given by the
Hitchin equations.
This implies that, as in the flat case, these drive a tunneling between the
different possible long string configurations of the orbifold $\sigma$-model
found at strong coupling.
The calculation of the WKB  expansion of the partition function goes exactly
like in the flat case as well as the evaluation of the gauge connection
determinants.

\paragraph{Conclusions and open questions}

It sounds very much like that matrix string theory should find its proper
place in gauge/string theory duality picture \cite{Polyakov}.
How and if it will happen is a nice challenging question we are facing
from this point of view.

\vspace{1 cm}

\noindent
{\bf Acknowlodgements}: 
I would like also to thank the organizers of the workshop 
``The quantum structure of spacetime and the geometric nature of fundamental
interactions'' in Leuven for the stimulating and nice atmosphere they have
created there.
Work supported by the European Community's Human Potential
Programme under contract HPRN-CT-2000-00131 Quantum Spacetime
in which G.B. is associated to Leuven.

\end{document}